\documentclass[prb,showpacs]{revtex4}% Physical Review B
\usepackage{graphicx}% Include figure files
\usepackage{graphics}% Include figure files
\usepackage{dcolumn}% Align table columns on decimal point
\usepackage{bm}% bold math
\usepackage[usenames]{color}
\begin{document}

\title{The roles of apex dipoles and field penetration in the physics of

charged, field emitting, single-walled carbon nanotubes}

\author{Jie Peng, Zhibing Li\footnote{Corresponding author:
stslzb@zsu.edu.cn}, Chunshan He, Guihua Chen, Weiliang Wang, Shaozhi
Deng, Ningsheng Xu\footnote{Corresponding author: stsxns@zsu.edu.cn}
}
\affiliation{The State Key Laboratory of Optoelectronic Materials
and Technologies \\ Department of Physics, Sun-Yat-Sen University,
Guangzhou, 510275, P.R. China}

\author{Xiao Zheng and GuanHua Chen}
\affiliation{Department of Chemistry, The University of Hong Kong,
Hong Kong, China}

\author{Chris J. Edgcombe}
\affiliation{PCS Group, Department of Physics, Cambridge University, UK}

\author{Richard G. Forbes}
\affiliation{Advanced Technology Institute, School of Electronics and Physical Sciences,
University of Surrey, Guildford, Surrey GU2 7XH, UK}

\date{ \today }

\vspace{1pt}

\begin{abstract}
A 1 $\mu$m long, field emitting, (5,5) single-walled, carbon nanotube (SWCNT) closed with a fullerene cap, and a similar open nanotube with hydrogen-atom termination, have been simulated using the MNDO (Modified Neglect of Diatomic Overlap) quantum-mechanical method. Both contain about  80 000 atoms. It is found that field penetration and band-bending, and various forms of chemically and electrically induced apex dipole, play roles. Field penetration may help to explain electroluminescence associated with field emitting carbon nanotubes. Charge-density oscillations, induced by the hydrogen adsorption, are also found. Many of the effects can be related to known effects that occur with metallic or semiconductor field emitters;  this helps both to explain the effects and to unify our knowledge about field electron emitters.  However, it is currently unclear how best to treat correlation-and-exchange effects when defining the CNT emission barrier. A new form of definition for the field enhancement factor (FEF) is used. Predicted FEF values for these SWCNTs are significantly less than values predicted by simple classical formulae. The FEF for the closed SWCNT decreases with applied field; the FEF for the H-terminated open SWCNT is less than the FEF for the closed SWCNT, but increases with applied field. Physical explanations for this behavior are proposed. Curved Fowler-Nordheim plots are predicted. Overall, the predicted field emission performance of the H-terminated open SWCNT is slightly better than that of the closed SWCNT, essentially because a C-H dipole is formed that reduces the height of the tunnelling barrier. In general, the physics of a charged SWCNT seems more complex than hitherto realised.
\end{abstract}

\pacs{73.22.-f, 73.21.-b, 73.63.Fg, 79.70.+q }% PACS, the Physics and Astronomy
                             % Classification Scheme.
\maketitle

\section{INTRODUCTION}

\subsection{General background}

The use of carbon nanotubes (CNTs) as field electron emitters has attracted many experimental and theoretical studies. This is mainly because CNTs have high aspect-ratios, which in turn means that they can be operated at low applied voltage. In the context of field electron emission (FE), the CNT is an interesting multi-scale system, in that there is a strong interaction between its mesoscopic length and the nanoscale details of electronic structure at the CNT apex. The electron emission is sensitive to the apex electronic structure, and thus needs to be treated quantum mechanically.

In the literature there are contradictory conclusions about the effect of hydrogen adsorption on the FE properties of carbon nanotubes. Zhou et al.\  \cite{ref1} and Kim et al.\  \cite{ref2,ref3} obtained the local density of states (LDOS) at the apex by \emph{ab initio} methods; they found that the LDOS at the charge-neutrality level was suppressed by the hydrogen. They therefore concluded that hydrogen adsorption reduces FE current density.  By contrast, Mayer et al. calculated the tip barrier using a dipole and point charge model~\cite{ref4}; they assumed that the tip barrier was reduced by the presence of the hydrogen, and concluded that hydrogen adsorption enhanced the FE current density. Mayer recently improved the model and illustrated the electrostatic potential around the carbon nanotube~\cite{ref5}. There is also confusion in experiments on the effect of hydrogen on FE from CNTs~\cite{ref6,ref7}. More extensive studies on this topic would obviously be useful.

A few years ago, it became possible to simulate long CNTs (up to several $\mu$m in length) by a hybrid approach involving both quantum mechanics and classical analysis~\cite{ref8,ref9}. A (5,5) single-walled carbon nanotube (SWCNT) closed with a fullerene hemisphere was simulated in this way (we refer to this as the ``closed CNT''). This work suggested that the superior emission characteristics of a CNT may involve, not only a large field enhancement factor, but also lowering of the tip barrier as result of field penetration. To investigate the effects of hydrogen adsorption, an open-ended (5,5) SWCNT, with the end terminated by hydrogen (H) atoms, has been simulated (we refer to this as the ``open CNT'').

Recently, we have developed our program to the point where a whole 1 $\mu$m long CNT, comprising about 80 000 atoms, can be simulated using only the quantum-mechanical part of the procedures described in Refs 8 and 9. Some small inconsistencies associated with the matching between the classical and quantum regions have thereby been removed. All results reported here have been regenerated in this way. We have also re-simulated the closed SWCNT with computational effort similar to that used for the open CNT. There have been no qualitatively significant changes in the outcomes as a result of this change in procedure.

In order to make theoretical comparisons, we also attempted to simulate a non-H-terminated open SWCNT.  This configuration is not expected to be ``real", because it implies the existence of unsaturated dangling bonds:  in practice, if such a structure came into existence, then the bonds would be expected to re-hybridise and the structure would be expected to reconstruct (most probably into a closed CNT).  In our simulations we can get results for the non-H-terminated open SWCNT for some applied fields.  But, as expected, it is not a stable system and sometimes the simulation results do not converge.  We do not have confidence that these results have any useful physical meaning, and do not present them.  We think that Mayer~\cite{ref5} was able to treat an open CNT successfully because his method of structure analysis was based on classical electrostatics rather than quantum mechanics.  We have thought it best to compare our analysis of the H-terminated open SWCNT with the equivalent analysis of a closed SWCNT.

It has been found that the open CNT differs from the closed one in several respects. The least expected results related to predicted field enhancement factors ($\gamma_s$ as defined here - see below). Application of the simple height/radius formula for a metal post with a hemispherical cap~\cite{ref10}, of apparently the same dimensions as our closed CNT (length 1 $\mu$m and geometrical radius 0.35 nm) estimates the field enhancement factor as around 2860; a better classical estimate, derived from eq.\ (20) in Ref.\ 10, is 1550.  (In these estimates, the ``geometrical radius" is defined by the positions of the carbon-atom nuclei.) Both our CNT configurations have $\gamma_s$ significantly less than this. The closed CNT has $\gamma_s$ in the range 750 to 1000, with $\gamma_s$ decreasing with increase in the applied macroscopic field $F_M$; the H-terminated open CNT has $\gamma_s$ in the range 370 to 450, with $\gamma_s$ increasing with increase in $F_M$.

In looking for the physical origin of these results, we re-examined wider aspects of the physics of charged CNTs, including the role of field penetration and possible electric-dipole effects near/at the CNT apex. It soon became clear that the physics of a charged CNT is more complex than hitherto realised. The aims of this paper are to present our numerical results, and set them in the context of reasoned hypotheses about the physics of charged CNTs.

\subsection{Theoretical background}

The normal FE convention is used that an electron extracting field is denoted by $F$ and is considered to be positive. This quantity $F$ is the negative of the quantity ``electric field'' defined in conventional electrostatics. The situation envisaged is that of a SWCNT standing upright on one plate of a parallel-plate arrangement, with the applied macroscopic field $F_M$ being the field between the plates, at a large distance from the SWCNT. The name ``barrier field" is used for the much higher local fields immediately above the CNT apex, that determine the shape of the electron tunnelling barrier.  Barrier field varies with position; more careful definitions of a field ($F_s$) of this type and of $\gamma_s$ are given later.

The conventional descriptions of FE and thermal electron emission, originally developed for metals, postulate the existence of an electron in the process of emission. This is sometimes called the ``external electron", but is here called the ``departing electron". The total potential governing its motion is historically called the ``motive"~\cite{ref11}. The corresponding total potential energy (PE) is called here the ``motive energy". The motive energy is what goes into the one-electron Schroedinger equation, and hence into approximate methods of solving it to obtain a tunnelling probability. Conventional theory identifies two components.

The first is the so-called ``electrostatic~\cite{ref11} component of electron potential energy" (EEPE). The EEPE relates to the potential that would be experienced by a (hypothetical) vanishingly small test charge moving in the field created by the \emph{total} emitter charge distribution, including the departing electron. (This may look illogical, but in real situations there is a high-voltage generator connected to the emitter, that pushes a replacement electron into the system as the departing electron leaves.) Thus, the EEPE is the PE that would apply to the motion of an \emph{extra} electron if the rest of the charge distribution in the metal or CNT did not react in any way to the presence of this electron.

The second component in the motive energy represents the system response to the departing electron. This is sometimes called the ``correlation-and-exchange'' (C\&E) component. For metals, the C\&E PE is almost always approximated (in the region where the tunnelling barrier exists) by a classical image PE. For a flat planar surface of a free-electron metal, this leads to the so-called Standard (Murphy-Good) Fowler-Nordheim-type (FN-type) Equation~\cite{ref12} for the FE current density. For a curved emitter surface, with a slightly different barrier shape, this leads to a FN-type equation with slightly different correction factors in the exponent and the pre-exponential~\cite{ref13,ref14}.

In principle, if the same general ideas apply to FE from CNTs as apply to metals, then the motive energy for a CNT should contain both ``electrostatic" and ``C\&E" components. However, there is continuing unpublished debate about whether, and how, a C\&E PE component should be included in the motive energy for FE from a CNT. No consensus exists on these issues, so in this paper we calculate only the electrostatic component (the EEPE). This is an useful first step, anyhow, (and was the strategy adopted by Fowler and Nordheim in their original work on FE from free-electron metals~\cite{ref15}). Qualitative effects associated with the SWCNT charge distribution should emerge correctly. Absolute values of predicted emission currents for a given applied field $F_M$ may be too low, but our comparisons of the emission behavior of closed and open CNTs should be qualitatively valid.

A brief description of the main calculation method is given in Section II. Section III then discusses the zero-field barrier shapes, and Section IV the results found when the applied macroscopic field is high enough to cause electron emission by tunnelling. Section V presents a more detailed discussion of field enhancement factors. Section VI considers estimates of emission current. Section VII discusses some specific issues, and Section VIII summarises our conclusions.

The following phrases are used with the technical meanings defined in Ref.\ 16. The $\emph{excess electron density}$ is the difference between the (simulated) actual electron density and the electron density that would exist if neutral atoms were brought together (in the absence of any applied field), with no change in their electron distributions. The $\emph{induced electron density}$ is the difference between the excess electron densities in the presence, and absence, of the applied field. The excess electron density in zero applied field represents chemically-induced changes. The induced electron density represents further changes induced by applying an electric field. The excess electron density in non-zero applied field includes both chemically induced and electrically induced effects. For diagrams, we use the convention that an increase in electron density is represented as a positive change in density (i.e.,  the convention is the reverse of that used in classical electrostatics).

\section{THE SIMULATION METHOD}

To simulate the SWCNT in realistic FE experimental conditions, a linear-scale algorithm is needed. A key decision is to divide the long CNT into sub-regions~\cite{ref8,ref9}. Each sub-region consists of 40 atoms, except the apex which consists of 80 atoms. A sub-region and its adjacent sub-region(s) form a subsystem (the dangling bonds are saturated by hydrogen atoms). Each subsystem was simulated by the ``Modified Neglect of Diatomic Overlap (MNDO)'' semi-empirical quantum-mechanical method~\cite{ref17}. The charges in other parts of the nanotube contribute interactions to the subsystem as if they were point charges. The principles of our general approach have been described elsewhere~\cite{ref8,ref9}. In the present paper, a total of 1600 to 2000 sub-regions is used in the quantum-mechanical part of the calculation.

A macroscopic field applied to the CNT would, if no electron redistribution occurred, create an electrostatic potential variation along the length of the CNT. In consequence, as in classical electrostatics, electrons are induced to move into the CNT from the substrate to which it is attached, in order to reduce the potential variation along the CNT. We shall assume that quasi-thermodynamic equilibrium exists, and that the electrochemical potential (Fermi level) is constant along the nanotube and equal to that of the substrate. In the semiconductor FE literature, e.g. Ref.\ 18, this is sometimes called the ``zero current approximation''.

As a result, the density (and induced density) of electrons increases along the CNT, with the highest induced densities at the CNT apex. To guarantee the constant Fermi level, the number of electrons in each sub-region must be adjusted so that it gives the desired Fermi level according to the Fermi-Dirac distribution. This is achieved by an iteration method.

The EEPE is denoted by $U$, and is the superposition of the following: (a) the PE due to the applied macroscopic field; (b) the PE due to the excess-charge distribution in the entire nanotube; and (c) the PE due to the related substrate surface charge distribution, the effect of which is represented by the images in the substrate of the excess CNT charges.

For positions close to atoms, in particular at the CNT apex, it is not sufficiently accurate to treat the electron distribution in terms of point charges. To obtain the apex surface PE barrier, the density matrices of the closest 7 sub-regions of the CNT were used. The induced charges in other sub-regions were treated as point charges, since they are less critical. To check the convergence, we used the density matrices of the closest 10 and closest 30 sub-regions in two test calculations, and found that results were not significantly different from those found using the closest 7 sub-regions.

\begin{figure}
\includegraphics[width=0.6\textwidth]{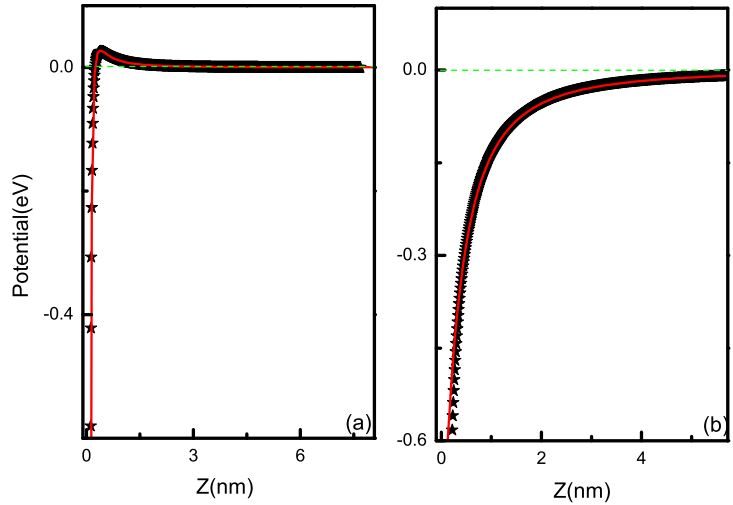}
\caption{\label{fig:1} (color figure) Total electrostatic electron potential energies (EEPEs) in the absence of applied field: (a) above the nucleus of a pentagonal-ring atom in the closed CNT; (b) above a hydrogen-atom nucleus in the H-terminated open CNT. The starred points and the solid curve show the calculated values and a fitted curve, respectively.}
\end{figure}

\section{INTRINSIC BARRIERS}

\subsection{Barrier shapes}

The motive energy in the vicinity of the CNT apex is important for FE.  When no applied field is present, this PE confines the electrons inside the CNT and provides the so-called intrinsic tip barrier. We denote its electrostatic component (the EEPE) by $U_0$. The zero of $U_0$ is taken as the EEPE of an electron at a point in remote space, in the absence of any applied field; it can also be identified as the local vacuum level just outside the metal substrate.

For the closed CNT, for $F_M = 0$, Fig.\ 1(a) shows how $U_0$ varies along a line, parallel to the central axis of the tube, that passes through the nucleus of an atom in the topmost pentagonal ring of carbon atoms; distance $z$ is measured from the atomic nucleus. The PE variation is, in fact, very similar along the axis itself. For the open CNT, Fig.\ 1(b) shows how $U_0$ varies along the wall of the nanotube and forwards into vacuum, along a line that runs parallel to the central axis of the tube, passing through one of the adsorbed hydrogen atoms; here, distance $z$ is measured from the hydrogen-atom nucleus.  For the closed CNT, the atoms in the top pentagonal ring are centred at 1009.00 nm from the substrate; for the open CNT, the hydrogen atoms are centred at 1008.69 nm from the substrate.

In Fig.\ 1(a), $U_0$ for the closed CNT has a small peak at a value of $z$ greater than zero. This form of EEPE variation is characteristic of that produced by a dipole  with its negative end outwards. One possibility is that this is the result of local asymmetry in the electron distribution in the fullerene hemisphere. Neutral flat graphene has no dipole or quadrupole moment normal to the plane of the graphene sheet. However, the CNT cap is highly curved, so a small spontaneous polarization might arise from the mixing of the p and s atomic states, thereby giving rise to a dipole and/or quadrupole moment.

Another possibility is that the effect is due (or partly due) to a small difference between the electronic structures of the CNT cylindrical wall and CNT quasi-hemispherical cap. There is both experimental and theoretical evidence for the existence of localized states at the end of closed CNTs~\cite{ref19,ref20,ref21,ref22}. A possibility is that the differences in local electronic structure are such that it is energetically favourable for electrons to move from the cylindrical CNT wall into a localised state associated with the top pentagonal ring (or into some other state associated with the cap), thus creating a dipole with its negative end outwards and raising the apex band-structure. This effect would resemble the upwards band-bending that occurs when electrons transfer from the surface region of a semiconductor to surface states in the band gap and below the bulk Fermi level~\cite{ref18}. In the on-line version of Fig.\ 5(b) in Ref.\ 9, the red spots on the left-hand-side of the diagram (that indicate a region of electron deficiency near the top of the CNT cylindrical wall) are possibly a signature of this effect.

For the open CNT, in the absence of the applied macroscopic field, local dipoles are formed at the H-terminated apex. This results from the electronegativity difference between hydrogen and carbon atoms. Since the electron tends to transfer from the hydrogen atom to the carbon atom, a dipole with its positive end outwards is formed. A dipole of this sense tends to reduce the surface electron barrier, as has been found in some experiments~\cite{ref23}.

To illustrate the charge transfer, we simulated the apex charge density of the H-terminated SWCNT in two ways: (1) we allowed charge to transfer from hydrogen to carbon, as it would in the real FE conditions; and (2) we artificially prevented the charge from transferring. The difference (1) minus (2) yields the excess electron density. Fig.\ 2 shows this density in a section through the central axis of the CNT. The CNT is not symmetrical about its central axis, so this section passes through a hydrogen atom only in the top half of Fig.\ 2(b). In the on-line version of the diagram, the deep blue region has lost electrons, the red region has gained electrons. The dipole associated with the C-H bond can be clearly seen.

Figure\ 2(b) also shows that, in our calculations, the field due to the C-H apex dipole has induced a charge-density oscillation in the neighbouring atoms, that decays with distance down the nanotube. We believe that, physically, a decaying charge-density oscillation is to be expected, for much the same reason that Friedel oscillations exist. Friedel oscillations in CNTs have been discussed by other authors, e.g. Ref.\ 24, though we are not aware of any previous suggestion that they would be caused by adsorbed hydrogen. However, our oscillations appear not to have the right wave-length for genuine Friedel oscillations, so it is possible that the detail in our results is a consequence of using the MNDO model. Nevertheless, we do believe that adsorbed hydrogen should induce charge-density oscillations and that the effect should be looked for in more sophisticated CNT models.

\begin{figure}
\includegraphics[width=0.6\textwidth]{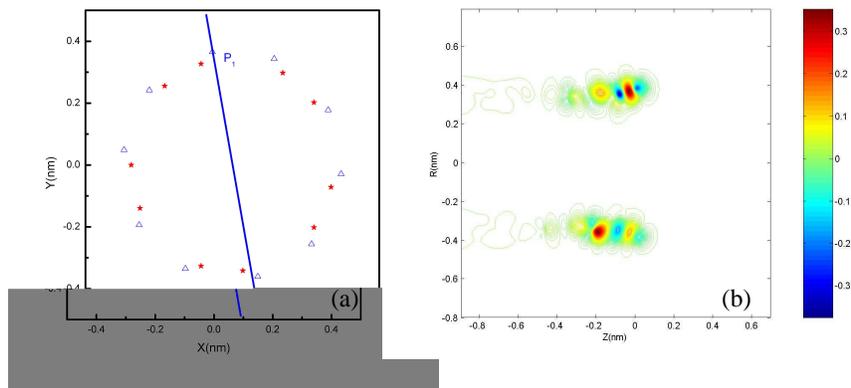}
\caption{\label{fig:2} (color figures) (a) Projection onto a plane transverse to the axis, showing the positions of the hydrogen atoms (blue $\bigtriangleup$) and the first ring of carbon atoms (red $\star$) at the end of an open-ended (5,5) SWCNT. The line shows the end view of the plane of Fig.\ 2(b). (b) The excess electron density at the apex of the hydrogen terminated CNT in the absence of the applied field, on a plane containing the axis of the CNT and passing through one of the adsorbed hydrogen atoms. The distances, from the substrate surface, of the hydrogen atom and its associated carbon atom are 1008.69 and 1008.60 nm, respectively. Positive density values indicate excess of electrons. The C-H dipole has its positive end outwards.}
\end{figure}

\subsection{A simple model}

We now present a simple model that brings out more clearly the difference between the closed and open CNTs. In Fig.\ 1, the deep potential well at $z=0$ is, in both cases, the screened coulomb PE due to the relevant atomic nucleus. For simplicity, we will attempt to model this part of the EEPE variation with a quadrupole PE term. We then add to this a dipole term, to represent the effects discussed above. We take both the dipole and the quadrupole to be located at the position $z=z_0$, on the line along which the barrier is considered. Thus we write

\begin{equation}
U_0(z)=-\frac{e}{4\pi\epsilon_0}\frac{P_0}{(z-z_0)^2}-\frac{e}{4\pi\epsilon_0}\frac{D_0}{(z-z_0)^3}
\end{equation}

\noindent
where $P_0$ and $D_0$ are, respectively, the strengths of the dipole and the quadrupole in the absence of the applied field. The convention is used that the dipole strength is positive if the dipole has its positive end outwards.

As shown, the two plots in  Fig.\ 1(a) and 1(b) are well fitted by the dipole-plus-quadrupole expression, and yield the following values: for the closed CNT: $z_0$= -0.035 nm, $P_0$ = -0.01 $e$ nm, $D_0$ = 0.0028 $e$ nm$^2$; for the H-terminated open CNT: $z_0$ = -0.7 nm, $P_0$ = 0.27 $e$ nm, $D_0$ = 0.014 $e$ nm$^2$. For the H-terminated open CNT the dipole is, as expected, of the opposite sign. This is the main result we wish to draw from this simple modelling; it confirms the impression given by the plots in Fig.\ 1.  A similar conclusion is drawn from a more complicated fitting model that uses a better mathematical representation of a screened coulomb potential.

For the open CNT, the values of all of $z_0$, $P_0$ and $D_0$ are much larger than for the closed CNT. We take this to be mainly a consequence of the charge-density oscillations exhibited in Fig.\ 2(b).

A conclusion, from the full simulations and from this model, is that for the closed CNT the apex dipole creates a field $F_{0,closed}$ in the same sense as the applied macroscopic field, whereas for the open CNT the apex dipole creates a field $F_{0,open}$ of the opposite sense to the applied macroscopic field. We return to this result below.

\section{EFFECTS OF THE APPLIED FIELD}

\subsection{Excess electron densities}

With the applied macroscopic field present, the overall electron density distribution for the closed CNT is similar to Fig.\ 5 of Ref.\ 9. Local-field induced dipoles (with their negative ends outwards) are clearly visible on the carbon atoms in the top pentagonal ring. The associated excess electron densities  are shown in more detail in Fig.\ 3 here.

A field-induced electric dipole layer always exists at the surface atoms of an electrically charged material (including metals).  A simple model based on a parallel-plate capacitor shows why. The positive charge on one electrode acts on the electrons of the surface atoms in the opposing plate. These electrons move away from their local atomic nuclei until the forces on their electrical centre due to these nuclei and to the distant  charge are equal and opposite. It follows that a dipole layer must exist~\cite{ref16}. However, the small closed CNT seems particularly favourable for strong dipoles, as depolarisation effects are smaller here than in many other contexts.

Surface dipoles with their negative ends outwards were first discussed in the context of FE by Drechsler, 50 years ago~\cite{ref25,ref26}, and are used by Mayer~\cite{ref5} in his analysis of open SWCNTs (without hydrogen termination), but we think ours has been the first detailed  quantum-mechanical model~\cite{ref9} to exhibit them clearly. The equivalent positive-end-outwards dipoles have long been part of the theory of the field ion emission techniques~\cite{ref16}, and  appear in advanced quantum-mechanical simulations~\cite{ref27} of positively-charged metal surfaces. The existence of these surface electric dipoles has significant implications in the physics of charged surfaces~\cite{ref16}.

\begin{figure}
\includegraphics[width=0.5\textwidth]{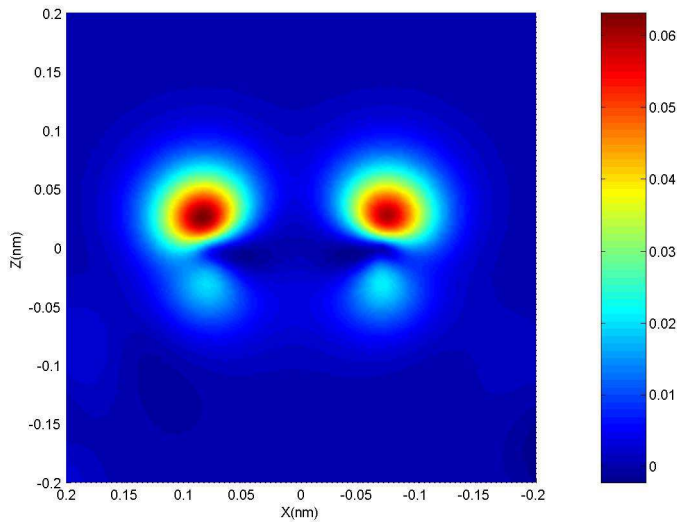}
\caption{\label{fig:3} (color figure) To illustrate the excess electron densities associated with two of the atoms in the top pentagonal ring of the closed CNT. Field-induced surface dipoles are prominently visible. The plane of the view pases through two of the carbon-atom nuclei and is parallel to (but does not include) the tip axis. The vertical axis represents distance parallel to the CNT axis.}
\end{figure}

\subsection{EEPE Distributions}

The overall EEPE distribution associated with the closed CNT is essentially similar to Fig.\ 3 in Ref.\ 8. The corresponding distribution associated with the open CNT is shown in Fig.\ 4 here. Fig.\ 4 and Fig.\ 2(a), between them, suggest that the field electron emission image associated with an H-terminated open CNT might take the form of a nearly circular ring.

Figure 5 compares the tip barriers for the closed and open CNTs in an applied macroscopic field of 11 V/$\mu$m. The upper curve is for the closed CNT, the lower one for the open CNT. For the closed CNT the horizontal axis has the same meaning as in Fig.\ 1(a), for the open-ended structure the same as in Fig.\ 1(b). The dashed line shows the Fermi level; for reasons of theoretical self-consistency associated with the code used in the calculations, the work-function of the SWCNT cylindrical wall has been taken as 5.08 eV.

\begin{figure}
\includegraphics[width=0.5\textwidth]{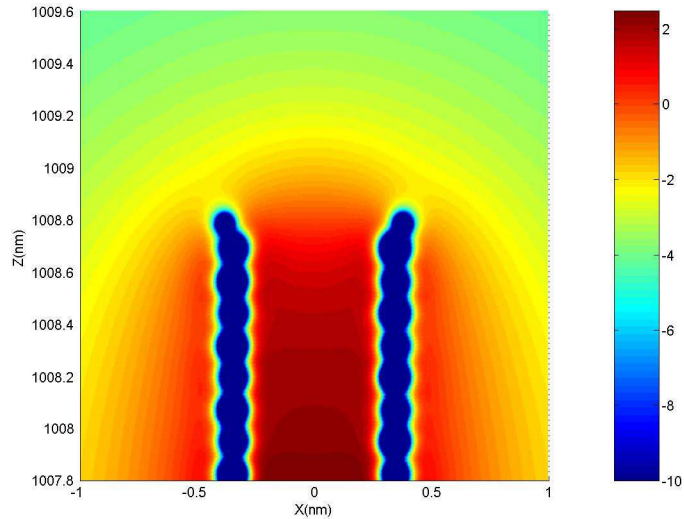}
\caption{\label{fig:4} (color figure) The electrostatic electron potential energy (EEPE) (in eV) in an intersection plane, parallel to the tube axis, through one of the terminated hydrogen atoms of the open CNT, in an applied macroscopic field of 10.0 V/$\mu$m.}
\end{figure}

\begin{figure}
\includegraphics[width=0.45\textwidth]{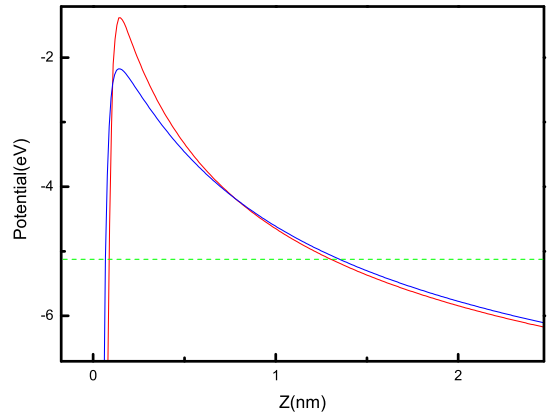}
\caption{\label{fig:5} (color figure) Variation of electrostatic electron potential energy for closed (red, upper curve) and open (blue, lower curve) CNTs, along the lines specified in the text, in an applied macroscopic field of 10.0 V/$\mu$m. The dotted green line represents the Fermi level, which is at -5.08 eV.}
\end{figure}

In making these comparisons we are assuming that, for the open CNT, the emission comes out of a covalent state associated with the C-H bonding, and that the transport characteristics of the CNTs are good enough to provide plenty of electrons into this state. In these circumstances, the barrier between the H atom and the vacuum is the main constraint on the electron emission.

Because the dipole field of the closed CNT is in the same direction as the applied macroscopic field, whereas the dipole field of the open CNT is in the opposite direction, the total calculated field for positive $z$-values will be greater for the closed CNT than for the open one. On the other hand, the dipole associated with the closed CNT will raise the apex surface potential barrier, whereas the C-H dipole of the open CNT will lower this barrier. Therefore, in applied fields, the tip barrier of the closed SWCNT is higher and sharper, while that for the open one is lower but smoother.

The barrier maximum is lowered by the applied field more for the open CNT than for the closed one. This is expected if the induced electron charge accumulates partly on the last carbon atom, rather than totally on the hydrogen atom.

We note that effects broadly similar to those reported here might be expected to occur with any adsorbate onto a carbon nanotube where the chemistry of the situation is such that: (1) a chemically-induced transfer of electron-type charge takes place inwards towards the carbon substrate, thereby creating an electric dipole with its positive end outwards ;  and (2) a covalent-type state is formed that spans both the adsorbate and some part of the carbon substrate, is at or closely below the CNT Fermi level, and is easily populated by charge transport from the main part of the CNT.

\section{FIELD ENHANCEMENT FACTORS}

\subsection{Results}

For a CNT, a quantitative characteristic of the tip barrier is its field enhancement factor. In the circumstances of these calculations it is difficult to define a quantity that corresponds exactly to the enhancement factor as defined in classical-conductor calculations. So, on the relevant line as specified earlier, we choose a fixed $z$-value, $z_s$, in space outside the peak in the barrier but close to the position where the field has its maximum absolute value (in the range of $F_M$ of interest). We then evaluate the field $F_s$ at position $z_s$, and define a field enhancement factor $\gamma_s$ as $F_s/F_M$.  In practice we take $z_s = 0.20$ nm for the closed CNT, $z_s = 0.24$ nm for the open CNT.

The variation of $\gamma_s$ with applied macroscopic field $F_M$ is shown in Figs 6(a) and 6(b). The values of $\gamma_s$ are significantly lower than simple classical-conductor calculations would suggest, and in neither case is the enhancement factor constant. For the closed CNT, $\gamma_s$ decreases linearly with $F_M$; for the open CNT, $\gamma_s$ increases with $F_M$, but the form of the variation is better described as a linear decrease with $1/{F_M}$.

\begin{figure}
\includegraphics[width=0.5\textwidth]{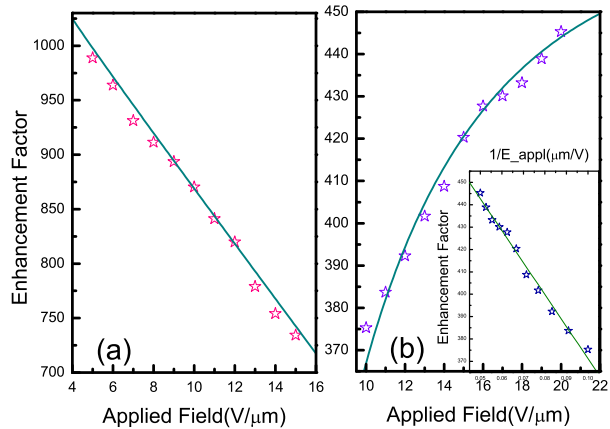}
\caption{\label{fig:6} (color figures) To show how the field enhancement  factor $\gamma_s$ varies with applied macroscopic field $F_M$ for (a) the closed CNT and (b) the  H-terminated open CNT. The solid lines are fitted curves. The inset in Fig.\ 6(b) shows the linear relationship between $\gamma_s$ and $1/{F_M}$.}
\end{figure}

\subsection{Theoretical discussion}

The simple classical-conductor calculations, noted earlier for the closed CNT, took the emitter radius to be defined in terms of the positions of the carbon-atom nuclei. In reality, the constant-potential surface assumed in the classical-conductor calculations needs to be taken at the outer boundary of the carbon atoms~\cite{ref16}, which makes the CNT electrical radius about 0.42 nm, and decreases the classical prediction of $\gamma_s$  to about 2380 for the simple formula and to about 1310 for eq.\ (20) in Ref.\ 10. Further, the position $z_s$ is outside this classical conductor surface by about 0.13 nm. If field fall-off outside the classical-conductor surface is taken into account, then an even lower classical $\gamma_s$-value (perhaps as low as 760) would be calculated. This is much the same as the values predicted for the closed CNT by the MNDO calculations. So it is unsurprising that the Fig.\ 6 values of $\gamma_s$ are significantly less than those derived from simple classical calculations. These lower values are also broadly consistent with the finding, by Bonard and colleagues~\cite{ref28}, that the simple ``height/radius" formula often overpredicted experimental $\gamma_s$-values by a factor of roughly two.

The $\gamma_s$-values calculated for a classical conductor would be constant, but ours are not. But, in fact, the enhancement factor as conventionally defined for an old-style metal field emitter is not strictly independent of field, either. Real field emitters, that expose facets with different values of local work-function, are surrounded by a system of so-called ``patch fields''~\cite{ref11,ref29}. Consider such an emitter mounted on one face of a parallel-plate arrangement. For such emitters, a simple theory of field enhancement exists. At a given part of the emitter surface, let the local field component induced by the applied voltage be $F_V$ and the component of patch field parallel to $F_V$  be  $F_P$, and suppose that these are independent. Let the total local field parallel to $F_V$ be $F_B$ and the macroscopic field be $F_M$. Then the total field enhancement factor $\gamma_B$ is given by

\begin{equation}
\gamma_B = (F_B/F_M)  =  (F_V+F_P)/F_M  =  \gamma_{class} + F_P/F_M
\end{equation}

\noindent
where $\gamma_{class}$ is the field enhancement factor you would get from an appropriate calculation involving a classical conductor of uniform work-function.

Since $F_P$ can be positive or negative, what one expects is that $\gamma_B$ varies with $1/{F_M}$, either positively or negatively. This is how the H-terminated open CNT behaves in our calculations. For the open CNT the dipole field $F_{0,open}$ discussed earlier is negative: thus a linear decrease in $\gamma_s$ as $1/{F_M}$ increases is expected (and observed).

But the closed CNT is behaving unexpectedly. Assume eq.\ (2) is applicable to a CNT, with $F_s$ taking the place of $F_B$, and $\gamma_s$ that of $\gamma_B$.  For $\gamma_s$ to decrease linearly as $F_M$ increases, there needs to be a source of local field $F_P$ that itself has a strength that varies as [$-{F_M}^2$], at least over the range of fields in question. We have examined several alternatives in search of a plausible physical explanation. So far, field penetration seems the only effect  likely to generate this form of dependence.

Physically, the argument for the existence of field reduction is as follows. If the nanotube were truly a metallic conductor, then the applied macroscopic field would generate an induced charge that is strongly localised near its apex. However, if the density of states of the carbon sheet is not high enough to allow sufficient charge to accumulate near the apex, then this induced charge will be spread over both the fullerene-type SWCNT cap and the near part of the cylindrical SWCNT column, and field penetration occurs.  Due to the post-like geometry, and the small size of the SWCNT under investigation, movement of induced charge away from the CNT apex will clearly have the effect of reducing the local field acting on the top atoms in the pentagonal ring, and this in turn would reduce the strength of the local field just outside the CNT apex.

Evidence that field penetration does occur in simulations of this kind can be seen in Fig.\ 4 of Ref.\ 8. A basic theoretical argument that field penetration might be expected to occur in the context of CNTs is as follows. If $S$ denotes the density of states per unit area, near the charge-neutrality level of a graphene or fullerene sheet, then the electron-type-charge per unit area $\sigma$ associated with downwards band-bending by an amount $|\Delta|$ is $\sigma=Se|\Delta|$. (The FE sign convention takes $\sigma$ as positive.) From Gauss's theorem the field close above this surface-charge density is $F =\sigma/\epsilon_0$. A value $S = 1$ eV$^{-1}$ nm$^{-2}$ may be derived from a formula given by Saito et al.~\cite{ref30}. Using this and the value $F = $ 10 V/$\mu$m yields $|\Delta| =  $0.5 eV. This gives some idea of the possible magnitude of such shifts. The effect is clearly sufficiently large that it needs to be taken into consideration. This is also the view of Chen et al.~\cite{ref31}

It is already known experimentally that  the field penetration into a field emitter does reduce the field above the emitter apex. From field ion microscope investigations long ago~\cite{ref32}, on an emitter made from uranium dioxide (which undergoes a change resembling a metal to semiconductor transition as temperature is reduced from about 110 K to about 80 K), it is known that field penetration into an old-style field emitter can cause reduction in the apex field  by a factor of order three. [What is actually observed is a three-fold increase in the so-called best image voltage as temperature is decreased through the above transition, which means that at the lower temperature you need to increase the applied voltage by a factor of three in order to get the same apex field as before.] The idea, that field penetration into a CNT field emitter might cause apex-field reductions of the size shown in Fig. 6(a), is thus not surprising.

Field penetration is known to be a phenomenon that can involve mathematical non-linearity~\cite{ref33}. So it seems not implausible that, in the lowest useful approximation, the effective field $F_a$ acting  on a pentagonal-ring atom should go approximately as:

\begin{equation}
F_a={c_1}{F_M}-{c_2}{F_M}^2
\end{equation}

\noindent
where  $c_1$ is a constant  and $c_2$ is a slowly varying (with $F_M$) parameter, with both $c_1$ and $c_2$ derived from a theory of field penetration for a CNT of the geometry under examination. The field $F_s$ (at position $z_s$, 0.20 nm away from a pentagonal-ring atom) would be expected to have the same dependence on $F_M$.  In the expression for the field enhancement factor $\gamma_s$, this would produce a component proportional to [$-F_M$], as observed in the MNDO calculations.

A variation in $\gamma_s$ that goes as $F_M$ would also occur if there were a contribution to the effective polarizability of a pentagonal-ring atom from a first-hyperpolarizability term that gave a contribution to its dipole strength of the form $(1/2)\beta {F_a}^2$. The first hyperpolarizability $\beta$ would be zero for a flat graphene sheet (or a spherically symmetric atom), but we think it possible that it could be non-zero for a carbon atom located in the sharply curved apex of a CNT. We find it difficult, however, to convince ourselves that this term would be negative rather than positive and of any significant size. If it exists, therefore, we think its most likely effect would be to reduce the consequences of field penetration slightly.

It might be asked why field penetration effects are, apparently, not found with the open CNT.  We think the answer is that, in this case, the chemically-induced C-H dipole is closer to the point at which $F_s$ is defined and plays the dominant role in determining how $F_s$ varies with $F_M$.

In summary, we currently think the most likely physical explanation, for the $\gamma_s$ dependence on $F_M$ observed in the MNDO calculations on the closed CNT, is that it is a side-effect of field penetration. We emphasize that this is field penetration into and partly down the column of the CNT, - in the context of a model that assumes that the field penetration causes band-bending but  the Fermi level stays constant. Classical-conductor arguments alone, if more carefully applied, would predict lower enhancement factor values than are found by unsophisticated application of the height/radius formula.  But they cannot predict a linear decrease with $F_M$.  If, however, part of the reason for the low $\gamma_s$ values is field penetration, then this dependence can be explained.

An increase of enhancement-factor with applied macroscopic field has been reported by Buldum and Lu~\cite{ref34}, in their classical-conductor-type calculations made as part of a simulation of emission from closed (5,5) SWCNTs very much shorter than ours (their lengths were 3.8 to 9 nm).  This variation appears to be the result of the way that they have defined their local field, $E_{loc}$, as the field at the point at which the electron leaves the barrier.  This point moves in towards the classical conductor surface as the applied macroscopic field increases, so an increase in field enhancement factor would be expected. This effect is different from those considered above; we have deliberately avoided the need to consider it, by choosing to define our local field $F_s$ at a fixed point in space, relative to the relevant atomic nucleus.

\section{EMISSION CURRENT}

For the closed and open CNTs, Fig.\ 7 shows the predicted emission currents $I$ calculated as described in Ref.\ 9, using the approximate formula

\begin{equation}
I = q_{ind}\nu D
\end{equation}

\noindent
Here, $D$ is the escape probability, calculated by carrying out a numerical integration on the relevant barrier, using the simple-JWKB (Jeffreys-Wentzel-Kramers-Brillouin) formula, $\nu$ is the frequency of electron collision with the barrier, and $q_{ind}$ is the total magnitude of the induced electron charge in the emitting region. Obviously, actual values used are field dependent. Typical values of $q_{ind}$ and $\nu$, for an applied macroscopic field of 12 V/$\mu$m, are as follows. For the closed CNT $q_{ind} = $0.16 $e$, $\nu = 1.2\times 10^{15}$ Hz; for the open CNT $q_{ind}= $0.45 $e$, $\nu = 1.7\times 10^{15}$ Hz. Figure 7 shows that the emission current from the open CNT is an order-of-magnitude greater than that from the closed one, for the same applied macroscopic field. This result goes in the same qualitative direction as the predictions of Mayer et al.~\cite{ref35} for a very much shorter SWCNT.

\begin{figure}
\includegraphics[width=0.5\textwidth]{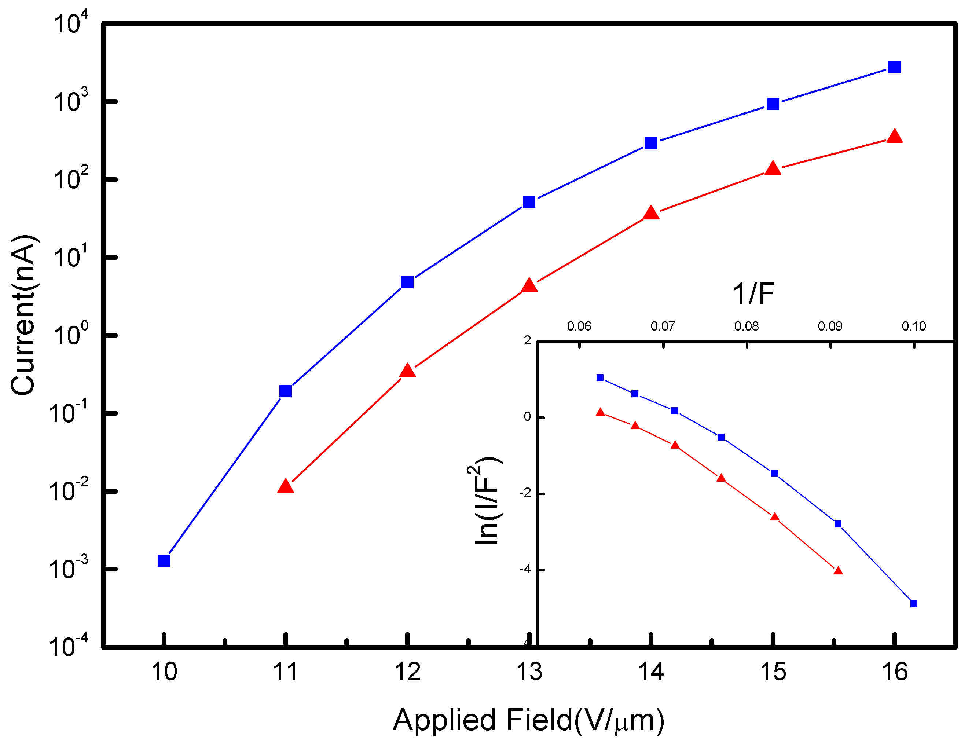}
\caption{\label{fig:7} (color figure) Emission current as a function of applied macroscopic field. The triangular points show the current from the closed CNT, the square points the current from the open CNT. The inset shows the relationship in the form of FN plots.}
\end{figure}

Currently, it is far from clear what the correct procedure is for calculating field-induced emission currents from a CNT, so we use a very simple formula.  Also, as noted earlier, correlation-and-exchange effects are neglected in this treatment.  The absolute emission-current values shown should be treated with great caution.  But we believe that qualitative comparisons between the closed and open CNT cases are valid.

The turn-on macroscopic field of the open CNT is slightly lower than that for the closed one. For this 1 $\mu$m long (5,5) SWCNT, using an onset current criterion of 1~nA, our simulations yield an onset macroscopic field of nearly 12.5 V/$\mu$m for the closed CNT, but 11.5 V/$\mu$m for the open CNT. The corresponding onset barrier-fields (values of field $F_s$) are 10 V/nm and 4.5 V/nm, respectively. The primary cause of these differences is that the hydrogen termination has produced a barrier-lowering effect.

An emitter of radius 0.42 nm has a cross-sectional area of about 0.55 nm$^2$, and a current of 1 nA drawn from this area corresponds to an average local current density of about $1.8\times 10^9$ A/m$^2$. For an emitter of work-function 5.08 eV, and the onset current criterion used here, the Elementary FN-type Equation predicts an onset barrier-field of 8.3 V/nm, and the Standard (Murphy-Good) FN-type Equation~\cite{ref12} an onset barrier-field of 5.7 V/nm. The difference between these two FN-type equations is that the latter includes an image PE in the potential barrier, whereas the former does not.

The apparently close agreement, for the closed CNT, between the barrier-field prediction (10 V/nm) of our calculations and the prediction (8.3 V/nm) of the Elementary FN-type Equation, must be largely a coincidence. This is because the barrier calculated here differs from the simple triangular barrier used in elementary FN theory, and there is no underlying theoretical reason to suppose that the electron supply to the CNT tunnelling barrier would be given by a Sommerfeld-type free-electron model. Nevertheless, the difference between the predictions of elementary and standard FN theory may give some idea of the reduction in predicted onset field to be expected if it were found correct to include a correlation-and-exchange component in the motive PE for electron emission for a CNT.

The inset to Fig.\ 7 shows that the predicted FN plots for both closed and open CNTs are curved downwards, rather than linear. The FN plots calculated by Buldum and Lu~\cite{ref34} for their short  closed (5,5) CNTs show much the same kind of curvature. Similar curvature has been found in FN plots derived from model calculations on classical spherical emitters, both in Ref.\ 13 and in unpublished work by one of us (RGF). Thus, we think that the smooth curvature seen in our FN plots and in some other FN plots is basically a classical electrostatic effect.  When an emitter is small and sharply curved, the field falls off quite rapidly with distance.  This makes the barrier shape different from that assumed in theory applicable to planar emitters, and leads to FN-plot curvature.

Experimental FN plots from CNTs do often exhibit convex (i.e., downwards) curvature. However, there are further possible explanations, such as ``saturation of the electron supply" (i.e. progressive breakdown of the assumption that the electron distribution inside the emitter, at its apex, is in thermodynamic equilibrium with electron pool in the substrate). If saturation occurs, then it will be at the high-current end of the range.  On the other hand, curvature of the cathode surface and equipotentials causes the whole of the FN plot to be slightly curved~\cite{ref36,ref37}. This is due to a change of emitting area, together with the slightly non-linear variation of the FN exponent with 1/(applied voltage).

For situations where FN plot curvature results from emitter sharpness, methods are being developed for determining many parameters of the cathode by using the curvature of the FN plot, together with the energy distribution of emitted current. However, measurement of the current-voltage relation with sufficient accuracy requires careful experimental procedure. At present it seems likely that, where emitter curvature is responsible for FNÐplot curvature, the latter will be a significant effect (and hence measurable) only for emitting regions with radius of a few nm or less.

\section{discussion}

\subsection{General issues}

These calculations, together with previous work~\cite{ref8,ref9}, have identified various physical effects that may influence field emission (FE) from single-walled carbon nanotubes (SWCNTs). They suggest that the physics of a charged nanotube may be significantly more complex than hitherto realised. In particular, we have established that field penetration should occur, and have identified four effects that might give rise to apex dipoles, namely: for the closed carbon nanotube (CNT) (1) mixing of orbitals at the carbon atoms at the highly curved fullerene cap; (2) transfer of electron charge between the near part of the CNT column and the cap, and (3) field-induced polarization of the apex carbon atoms; for the open CNT (4) electron transfer associated with hydrogen adsorption.  It has also been shown, for the open CNT, that hydrogen termination leads to charge-density oscillations.

The existence of field penetration might help to explain the electroluminescence reported by Bonard et al.~\cite{ref38}. Field penetration would allow both states in their two-state mechanism to be ``further up the band-structure" (and hence closer to the vacuum level) than they assumed. In consequence, the lower state need not be quite such a deep level, and this would make it easier for a hole to be created in such a state by tunnelling.

The CNT end-structure has an important influence on the tip barrier, and hence on FE current. For the open CNT, the apex dipole resulting from hydrogen termination plays a dominant role, in that it reduces the height of the tunnelling barrier. In zero applied field the electrostatic component of electron potential energy (EEPE) variation is relatively sharp for the closed CNT, but smoother and longer range for the open CNT, due to this large C-H dipole.

However, Fig.\ 1 shows that in both cases, as $z$ gets larger, the EEPE climbs back to the level of the reference zero within about 5 nm of the last CNT carbon atom. This behavior is markedly different from that associated with hydrogen adsorption on a flat metal plane (assuming the rest of the emitter is clean): in the metal-plane case there is a potential energy (PE) plateau just outside the surface, and the EEPE climbs back to the reference level over a distance comparable with the diameter of the plane. The absence of a PE plateau in the case of a H-terminated CNT makes it difficult to create a meaningful definition of local work-function that can be used in the presence of an applied field, and this in turn makes it difficult to estimate tunnelling probability by thinking of the hydrogen adsorption as leading to a reduction in work-function and then applying some simple formula.

Nevertheless, barrier reduction due to the C-H dipole is effective. Figure 2 confirms that, with the applied field present, the barrier height is lower for the open CNT. For a given applied field, the emission current is higher for the open CNT than for the closed CNT, even though the field enhancement factor (as calculated here) for the open CNT is significantly smaller. The turn-on macroscopic field is also lower for the open CNT.

Figure 2 shows that one expects the field $F_s$ to be lower for the open CNT than for the closed CNT, as we have found. Figure 6 shows that the field enhancement factor $\gamma_s$ for the open (H-terminated) CNT is about half that for the closed CNT. This may seem counter-intuitive, because in a certain sense the H-terminated open CNT looks sharper than the closed CNT. The reason for the effect is that the C-H dipole field opposes the applied macroscopic field.

We have found that the H-terminated open CNT should have slightly the better field emission performance. This finding does not necessarily contradict previous \emph{ab initio} calculations~\cite{ref1,ref2}. This is because, in our picture, electrons are emitted from energy states above the charge-neutrality level; in this case the density of states in the immediate vicinity of the charge-neutrality level may be of reduced relevance. As already noted, an aspect of this better performance is that the open CNT turns on at a slightly lower applied macroscopic field (11.5 V/$\mu$m, as opposed to 12.5 V/$\mu$m). However, this difference is slight. So, in the choice between SWCNT forms (assuming both could be manufactured reliably), other technological factors such as system stability would probably be of more significance.

\subsection{The correlation-and-exchange problem}

As already noted, a possible deficiency is that we neglect correlation-and-exchange (C\&E) effects associated with the departing electron. We have thought it useful to look at the electrostatic effects by themselves, first. We also wish to draw wider attention to apparent difficulties in deciding whether, and how, to take C\&E effects into account.

For FE from metals, the C\&E component in the motive energy is the metal response to a departing $\emph{point}$ electron: in the first approximation this is the well-known image PE, but there are more-detailed models involving interaction with a C\&E hole~\cite{ref39}, or involving some form of density functional theory. On the other hand, in the calculation of electronic ground states, every electron is represented by a delocalized wave-function, and the final result is a self-consistent solution in which exchange effects have been taken into account. A central question is as follows. In determining the C\&E response of the CNT to the presence of an electron in the process of emission (which we need to do, in order to create the motive-energy expression that goes into the one-electron Schroedinger equation), what shall we assume about the spatial distribution of the electron being emitted?

The conventional approach is to treat the departing electron as a point for the purposes of determining the system response, but treat it as delocalized and wave-like for the purpose of determining tunnelling probability). It would be helpful to have access to a clear explicit proof that this is the correct fundamental way to apply quantum mechanics.  We offer this issue for wider discussion.

\subsection{Use of the MNDO method}

Exact quantum-mechanical analysis of the surfaces of charged systems is notoriously difficult, especially in situations of limited symmetry. This is because (as the surface dipoles discussed earlier show) the charge is associated with a strong modification of the system wave-functions in the vicinity of the topmost layer of atoms (as compared with the corresponding uncharged surface). The MNDO method used here is a longstanding but relatively unsophisticated method, of limited numerical accuracy. Its advantage, in the analysis of charged surfaces, is that it is likely to bring out any special local atomic-level phenomena that may physically exist. It is not surprising that our results exhibit several such phenomena.  At a qualitative and semi-quantitative level, most of these are clearly consistent with experimental evidence and/or more general theoretical arguments. The main value of the present work is that it suggests and provides support to hypotheses that these effects do exist physically. It also provides some provisional numerical estimates. These effects should be looked for in more sophisticated treatments capable of greater numerical accuracy.

\section{ CONCLUSIONS}

In summary, this work (and our previous papers) has confirmed that, when calculating surface potential energy barriers and field enhancement factors, it is not satisfactory to treat single-walled carbon nanotubes as if they were solid metal objects. Essential differences are that: (1) the density of states near and above the charge neutrality level is much lower for a single-walled carbon nanotube than it is for a metal, so field penetration and band-bending effects may occur, especially near the emitter apex; and (2) various local-dipole effects may occur near/at the emitter apex. Another difference, not brought out in this paper, is that electron energy is effectively quantised in the direction normal to the fullerene-type carbon sheet that forms the emitting cap; this will affect observed energy distributions~\cite{ref38}.

The hydrogen-terminated open single-walled carbon nanotube behaves differently from the closed tube, the main differences being the variation of field enhancement factor with applied macroscopic field (increases with $F_M$ rather than decreases), the presence of hydrogen-induced charge-density oscillations, and the slightly better emission performance (associated with barrier reduction by the C-H dipole).

In some respects, field emission from a carbon nanotube (even from a so-called ``metallic" CNT) is more analogous to field emission from a semiconductor than to field emission from a metal, with the best simple analogy perhaps being to a situation where the conduction band is significantly degenerate and the resulting triangular well has quantised levels in it. But really the carbon nanotube needs to be treated as a field emission situation in its own right. There is, of course, no reason to expect that its behavior will be well described by the Standard Fowler-Nordheim-type Equation or by the simpler FN-type equations sometimes used in the literature.

There are two important messages from this work. First, that the physics of charged nanotubes seems, even qualitatively, to be more complex than has hitherto been realised. Second, that several effects uncovered here have equivalents in conventional field emission from larger metal and/or semiconductor emitters: so studies of this kind may help unify knowledge concerning field electron emission.

\section{ AKNOWLEDGEMENTS }

Ningsheng Xu, Zhibing Li and Shaozhi Deng gratefully acknowledge the financial support of the project from the National Natural Science Foundation of China (Grant No. 50021202, 50329201, U0634002,60571035 90103028, 90306016, and 10674182), Science and Technology Ministry of China (Grant No. 2003CB314701), Education Ministry of China, the Science and Technology Department of Guangdong Province (Grant No.03039) and the Science and Technology Department of Guangzhou City. Xiao Zheng and Guan-Hua Chen acknowledge gratefully the support from the Hong Kong Research Grant Council (RGC HKU7012/04P) and the Committee for Research and Conference Grants (CRCG) of the University of Hong Kong.

\clearpage


\vspace{1pt}



\begin{thebibliography}{99}

\bibitem{ref1} G. Zhou, W. H. Duan, and  B. G. Gu, Phys. Rev. Lett. \textbf{87}, 095504 (2001).

\bibitem{ref2} C. Kim, Y. S. Choi, S. M. Lee, J. T. Park, B. Kim, and Y. H. Lee, J. Am. Chem. Soc. \textbf{124}, 9906 (2002).

\bibitem{ref3} C. Kim, K. Y. Seo, B. Kim, N. Park, Y. S. Choi, K. A. Park, and Y. H. Lee et al, Phys. Rev. B \textbf{68}, 115403 (2003).

\bibitem{ref4} A. Mayer, N. M. Miskovsky, P. H. Cutler, and Ph. Lambin, Phys. Rev. B \textbf{68}, 235401 (2003).

\bibitem{ref5} A. Mayer, Phys. Rev. B \textbf{71}, 235333 (2005).

\bibitem{ref6} C. Y. Zhi, X. D. Bai and E. G. Wang, Appl. Phys. Lett. \textbf{81}, 1690 (2002).

\bibitem{ref7} A. Wadhawan et al., Appl. Phys. Lett. \textbf{79}, 1867 (2001).

\bibitem{ref8} X. Zheng, G. H. Chen, Z.B. Li, S. Deng, and N. Xu, Phys. Rev. Lett. \textbf{92}, 106803 (2004).

\bibitem{ref9} J. Peng, Z. B.  Li, C. S. He, X. Z. Deng, N. S. Xu, X. Zheng, and G. H.  Chen, Phys. Rev. B \textbf{72}, 235106 (2005).

\bibitem{ref10} R. G. Forbes, C. J. Edgcombe, and U. Valdr$\grave{\rm{e}}$, Ultramicroscopy \textbf{95}, 57 (2003).

\bibitem{ref11} C. Herring and M. H. Nichols, Rev. Mod. Phys. \textbf{21}, 185 (1949).

\bibitem{ref12} E. L. Murphy and R. H. Good, Phys. Rev. textbf{102}, 1464 (1956).

\bibitem{ref13} C. J. Edgcombe and A. M. Johansen, J, Vac. Sci. Technol. B \textbf{21}, 1519 (2003).

\bibitem{ref14} R. G. Forbes, Surf. Interface Anal.  \textbf{36}, 395 (2004).

\bibitem{ref15} R. H. Fowler and L. W. Nordheim, Proc. Roy. Soc. London, Ser. A \textbf{119}, 173 (1928).

\bibitem{ref16} R. G. Forbes, Ultramicroscopy \textbf{95}, 1 (2003).

\bibitem{ref17} M. J. S. Dewar and W. Thiel, J. Amer. Chem. Soc. \textbf{99}, 4899 (1977).

\bibitem{ref18} A. Modinos, \emph{Field, Thermionic, and Secondary ElectronEmission Spectroscopy }(Plenum Publishing Co.,New York, 1984).

\bibitem{ref19} R. Tamura and M. Tsukada, Phys. Rev. B \textbf{52}, 6015 (1995).

\bibitem{ref20} D. L. Carroll, P. Redlich, P. M. Ajayan, J. C. Charlier, X. Blase, A. De Vita, and R. Car, Phys. Rev. Lett. \textbf{78}, 2811 (1997).

\bibitem{ref21} P. Kim, T. W. Odom, J.-L. Huang, and C. M. Lieber, Phys. Rev. Lett. \textbf{82}, 1225 (1999).

\bibitem{ref22} S. Han and J. Ihm, Phys. Rev. B \textbf{61}, 9986 (2000).

\bibitem{ref23} P. Ruffieux, O. Groning, M. Bielmann, P. Mauron, L. Schlapbach, and P. Groening, Phys. Rev. B \textbf{66}, 245416 (2002).

\bibitem{ref24} M. F. Lin and D. S. Chuu, Phys. Rev. B \textbf{56}, 4996 (1997).

\bibitem{ref25} M. Drechsler, Z. Elektrochem. \textbf{61}, 48 (1957).

\bibitem{ref26} J.A. Becker, Study of Surfaces by Using New Tools.  In:  F. Seitz and D. Turnbull (eds.), Solid State Physics: Advances in Research and Applications, Vol. 7 (Academic, New York. 1958).

\bibitem{ref27} G. C. Aers and J. E. Inglesfield, Surface Sci. \textbf{217}, 367 (1989).

\bibitem{ref28} J.-M. Bonard, private communication to CJE and RGF, October 2002.

\bibitem{ref29} L. K. Hansen, J. Appl. Phys. \textbf{37}, 4498 (1966).

\bibitem{ref30} R. Saito, G. Dresselhaus and M. S. Dresselhaus, Physical Properties of Carbon Nanotubes (Imperial College Press, London, 1998).

\bibitem{ref31} C. W. Chen, M. H. Lee and S. J. Clark, Appl. Surface Sci. \textbf{228}, 143 (2004).

\bibitem{ref32} R. Morgan, J. Materials Sci. \textbf{5}, 445 (1970).

\bibitem{ref33} W. Moench, Semiconductor Surfaces and Interfaces (Springer, Berlin, 1995).

\bibitem{ref34} A. Buldum and J. P. Lu,  Phys. Rev. Lett. \textbf{91}, 236801 (1999).

\bibitem{ref35} A. Mayer, N. M. Miskovsky, and P. H. Cutler, Appl. Phys. Lett. \textbf{79}, 3338 (2001).

\bibitem{ref36} C. J. Edgcombe and N. de Jonge, J. Vac. Sci. Tech. B \textbf{24}, 869 (2006).

\bibitem{ref37} R. G. Forbes and J. H. B. Deane, in: Technical Digest, 18th Intern. Vacuum Nanoelectronics Conf., Oxford, July 2005, (ISBN: 0-7803-8397-4) (Eds: S.E. Huq, P.R. Wilshaw) (IEEE, Piscataway, NJ, 2005), p. 109.

\bibitem{ref38} J.-M. Bonard, J.-P. Salvetat, T. Stoeckli, L. Forro, and A. Chatelain, Appl. Phys. A. \textbf{69}, 1245 (1999).

\bibitem{ref39} T. L. Loucks and P. H. Cutler, J. Phys. Chem. Sol. \textbf{25}, 105 (1964).



\end{thebibliography}
\end{document}